\documentclass[traditabstract]{aa} % for the letters 
\usepackage{txfonts}
\usepackage{graphicx}

\begin{document}

\title{On the naming convention used for multiple star systems and extrasolar planets}
% \subtitle{A call for discussion}

\author{
F. V. Hessman\inst{1}
\and V. S. Dhillon\inst{2}
\and D. E. Winget\inst{3}
\and M. R. Schreiber\inst{4}
\and K. Horne\inst{5}
\and T. R. Marsh\inst{6}
\and E. Guenther\inst{7}
\and A. Schwope\inst{8}
\and U. Heber\inst{9}
}

\institute{
Institut f\"ur Astrophysik, Georg-August-Universit\"at, Friedrich-Hund-Platz 1, 37077, G\"ottingen, Germany
% \email{hessman@astro.physik.uni-goettingen.de}
\and
Dept. of Physics \& Astronomy, University of Sheffield, Sheffield S3 7RH, U.K.
\and
Dept. of Astronomy,  University of Texas at Austin, RLM 16.236, Austin, TX 78712, USA
\and
Departamento de Fisica y Astronomia, Universidad de Valparaiso, Av. Gran Bretana 1111, Valparaiso, Chile
\and
SUPA Physics \& Astronomy, University of St. Andrews, KY16 9SS, St. Andrews, Scotland
\and
Dept. of Physics \& Astronomy, University of Warwick, Coventry, CV4 7AL, U.K.
\and
Th\"uringer Landessternwarte, Sternwarte 5, 7778 Tautenburg, Germany
\and
Astrophysikalisches Institut Potsdam, An der Sternwarte 16, 14482 Potsdam, Germany
\and
Dr. Remeis-Sternwarte, Astronomisches Institut der Universit\"at Erlangen-N\"urnberg, Sternwartstr. 7, 96049 Bamberg, Germany
}

% \date{Received September, \ldots 2010 / accepted \ldots 2010}

\authorrunning{Hessman et al.} 
\titlerunning{Exoplanet Naming Convention}

% Abstract: Context, Aims, Methods, Results, [Conclusions]
\abstract { % Context
The present naming convention for extrasolar planets used by the vast majority of researchers in the field is based upon an interpretation of the provisional I.A.U. standard for multiple star systems.
With the existence of hundreds of exoplanets around single stars named by this convention and a handful of exoplanets around binary stars -- circumbinary planets --
it has become necessary to find a uniform and useful naming convention for the latter which is maximally compatible with the single host-star convention and which captures as much of the dynamical information about the planet as possible.
We propose a simple and generic naming convention for all exoplanets which follows the provisional I.A.U. standard but more clearly indicates their dynamical status.
The proposed convention is compatible with present usage and easily extendible to exoplanets around stars in systems of arbitrary multiplicity.
We invite comments and discussion on the proposed convention, in the hope of a timely adoption by the I.A.U. Commissions 5, 8+24, 26, 42, 45 and 53.
}
{ % Aims 
% With the existence of hundreds of exoplanets around single stars named by this convention and a handful of exoplanets around binary stars -- circumbinary planets --it has become necessary to find a uniform and useful naming convention for the latter which is maximally compatible with the single host-star convention and which captures as much of the dynamical information about the planet as possible.
}
{ % Methods
% We propose a simple and generic naming convention for all exoplanets which follows the provisional I.A.U. standard but more clearly indicates their dynamical status.
}
{ % Results
% The proposed convention is compatible with present usage and easily extendible to exoplanets around stars in systems of arbitrary multiplicity.
}
{ % Conclusions
% We invite comments and discussion on the proposed convention, in the hope of a timely adoption by the I.A.U. Commissions 5, 8+24, 26, 42, 45 and 53.
}

\maketitle

\section{Introduction}

When the first extrasolar planets (``exoplanets'') were discovered, it became necessary to find names for individual objects.
Given the large numbers of exoplanets, the Solar System convention of using mythological or other real names was utterly impractical and astronomically uninformative.

There already exist several naming conventions which have grown out of the historical needs of the visual and spectroscopic binary communities.
For instance, the components of visual binaries tend to be labeled with capital letters (e.g. $\xi$\,UMa\,A \& B), whereas spectroscopic binaries tend to be labeled with small-case letters or numbers (e.g. $\xi$\,UMa\,A consists of two stars, ``Aa'' and ``Ab'', ``Aa'' being the primary).
However, this system is neither officially defined and sanctioned by the I.A.U. nor is its use in the literature uniform.
Indeed, the notation used in earlier literature was often the opposite convention, i.e. capital letters for primary stars and the matching lower-case letters for the secondaries.
For example, the names of the components of $\xi$\,UMa used by Aitken (\cite{aitken}; p. 249),  `A'' \& ''a'' for the two components of the ``A'' system and ``B'' \& ``b'' for the two components of the ``B'' system, are carefully renamed by Griffin (\cite{griffin}; p. 276) to ``Aa'', ``Ab'', ``Ba'', and ``Bb''.
Most authors relieve themselves from the obviously onerous task of giving the components of spectroscopic or astrometric binaries names by using the terms ``primary'' and ``secondary'' or designating them as "1" or "2", particularly as indices of dynamical parameters.

This problem with the naming of stars in multiple systems is well-known and the source of many discussions in various commissions within the I.A.U.  (see Hartkopf \& Mason \cite{HandM}).
The provisional working standard adopted during the XXIV I.A.U. convention is that of the Washington Mulitplicity Catalog (WMC), which uses the following system:
\begin{itemize}
\item the brightest component is called ``A'', whether it is initially resolved into sub-components or not;
\item subsequent distinct components not contained within ``A'' are labeled ``B'', ``C'', etc.;
\item sub-components are designated by the concatenation of one or more suffixes with the primary label, starting with lower-case letters for the 2nd hierarchical level and then with numbers for the 3rd.
\end{itemize}
This system makes no distinction between stellar, sub-stellar, and planetary objects but does express a clear hierarchical structure.
One problem with this system is that the discovery hierarchy is not necessarily identical with the dynamical hierarchy: does HD\,97950\,C orbit around HD\,97950\.A or perhaps around HD\,97950\,A+HD\,97950\,B?
Systems of the latter type are often expressed as HD\,97950\,AB, i.e. by concatenating the component suffixes: the WMC contains references to things like ``A-BC'', i.e. a triple system consisting of the brightest component, ``A'', orbiting around a fainter binary, ``B''+``C''.
However, this nomenclature is also not adequate enough to express the dynamical state of just one of the components.
Another problem with the WMC nomenclature is that the names are purely accidental and/or historical: we are used to referring to ``Sirius B'', not ``Sirius Ab''; and ``Sirius AB'' can be a reference to both stars together or a capitalized misprint of just one.
Finally, the decision, what to call the ``A'' component is arbitrary: historically, visual binaries yielded the widest binaries and so defined the upper-case usage and spectroscopic binaries came later, inducing the lower-case additions, but it is equally possible to first discover a binary using some other method, which then implies ``A'' and ``B'' (instead of ``Aa'' and ``Ab'') components, and to discover a companion system later astrometrically, which then would be the ``C'' component (rather than ``B''), even though the latter might have a totally different dynamical relationship to the first two objects.

Given that the situation for multiple stars is confusing enough, when it came to naming exoplanets the simplest solution was to name the planets around single stars using a variation of the WMC convention: if the host planetary system is the ``A'' component (i.e. may or may not be a member of a hierarchical stellar system), then the first exoplanet was considered to be the secondary sub-component and should have been given the suffix ``Ab''.
For example, 51\,Peg\,Aa is the host star in the planetary system 51\,Peg, and the first exoplanet is then 51\,Peg\,Ab.
Since most exoplanets are in single star systems, the implicit ``A'' designation was simply dropped, leaving the exoplanet name with the lower-case letter only: 51\,Peg\,b.
This meant that researchers from the exoplanetary community have adopted what we will refer to as the ``lower-case b'' nomenclature, i.e. without the reference to the primary component and probably have little or no knowledge of the historical nomenclature behind it.
The usage of the lower-case b notation is not universal, however:  e.g. the planets around the pulsar PSR\,1257+12 were long labeled numerically starting with the index 1 but have also been labeled with lower-case letters starting with ``a'' (Currie \& Hansen \cite{currie}) and upper-case letters starting with ``A'' (Wolszczan \cite{wolszczan}).
Thus, the situation is far from uniform even for exoplanets.
The usual notation becomes dangerous when considering exoplanets around the stars in binary systems, e.g. $\tau$\,Boo\,b is the name given to the first planet discovered around the primary star of the $\tau$\,Boo binary system, but could $\tau$\,Boo\,c be the 2nd star around the primary or the 1st star around the secondary?
Fortunately, now that there are a few planets of this kind -- 16\,Cyg\,Bb, 30\,Ari\,Bb, $\tau$\,Boo\,Ab, HD\,178911\,Bb, HD\,41004\,Ab \& Bb -- the planets around the secondary stars have to date been correctly named.

The implicit system for exoplanet names utterly failed with the discovery of circumbinary planets in systems like HW\,Vir (2 planets; Lee et al. 2009), DP\,Leo (1 planet; Qian et al. 2010), and NN\,Ser (2 planets; Beuermann et al. 2010).
Lee et al. tried to circumvent the naming problem in HW\,Vir by calling the two planets ``HW Vir 3'' and ``HW Vir 4'', i.e. the latter is the 4th object -- stellar or planetary -- discovered in the system HW\,Vir, which is inconsistent with a similar convention already used for pulsar planets in the literature, where the first planet was labelled, e.g. PSR\,1257+12\,\#1 (these pulsar exoplanets are now registered in exoplanet.eu using the lower-case b notation).
In the case of NN\,Ser, Beuermann et al. were confronted with multiple suggestions from various offical sources and finally chose to use the designation NN\,Ser\,c and NN\,Ser\,d, i.e. implicitly NN\,Ser\,Ac and NN\,Ser\,Ad with the central very close binary system composed of NN\,Ser\,Aa and NN\,Ser\,Ab.
This solution conflicts with the standard usage of ``A'' and ``B'' for the primary and secondary stars in (pre-)cataclysmic variables and places the two stars and the two planets on the same hierarchical level.
The official alternative would have been either to declare NN\,Ser\,Aa+Ab as one dynamical component with the exoplanets NN\,Ser\,B and NN\,Ser\,C orbiting around it,
which would have described the dynamical separation of the stars and planets more explicitly but would have placed the planets on a higher hierarchical level than the stars (at least semantically) or
to adopt the standard usage of NN\,Ser\,A \& B for the close binary stars, leaving the planets as NN\,Ser\,C \& D.
No matter how hard one tries, the designations for the two circumbinary planetary systems are confusing and seemingly incompatible with the common usage for the other exoplanets.

Naming conventions are not physically important -- no one really cares if an object is called Sirius\,B, $\alpha$ CMa\,Ab, GJ\,244 \#2, RXF\,J064508.6-164240, or ``Rover'', but names convey both historical {\em and} physical information about the object and the naming convention used should at least not confuse.
This is particularly true for the benefit of observers, who are definitely interested in knowing which object on the sky is meant by what name.
Unlike the multiple star community, which is suffering from over a century of jumbled naming conventions, the exoplanet community is still sufficiently young that it is possible to adopt a uniform nomenclature which maximizes the usefulness of the names and minimizes the amount of confusion while consciously staying as close as possible to the provisional I.A.U. multiple star naming standard.
The purpose of this letter is to propose a simple, maximally compatible and yet physically informative solution for this problem, in the hopes that the I.A.U. (or at least Commision 53) would eventually adopt it for universal usage.

\section{A simple proposal}

We propose the following exoplanet naming convention that preserves as much of the present names as possible but is flexible enough to be used in any planetary configuration.
\begin{description}
% 1
\item[Rule 1.] {\em The formal name of an exoplanet is obtained by appending the appropriate suffixes to the formal name of the host star or stellar system.  The upper hierarchy is defined by upper-case letters, followed by lower-case letters, followed by numbers, etc.  The naming order within a hierarchical level is for the order of discovery only.} \\[0.1ex]
This rule corresponds to the present provisional I.A.U.-sanctioned WMC naming convention, including the present use, e.g., of the ``Bb'' notation for the exoplanets around the secondaries in binaries.\\
% 2
\item [Rule 2.] {\em Whenever the leading capital letter designation is missing, this is interpreted as being an informal form with an implicit ``A'' unless otherwise explicitly stated.} \\[0.1ex]
This rule corresponds to the present exoplanet community usage for planets around single stars (e.g. 51\,Peg\,b $\equiv$ 51\,Peg\,Ab).    Thus, all of the present names for 99\% of the planets around single stars are preserved as informal forms of the I.A.U. sanctioned provisional standard .\\
% 3
\item [Rule 3.] {\em As an alternative to the nomenclature standard in rule\#1, a hierarchical relationship can be expressed by concatenating the names of the higher order system and placing them in parentheses, after which the suffix for a lower order system is added.} \\[0.1ex]
This rule permits one to keep the lower-case b notation even when the previous hierarchical naming would suggest the use of a different suffix.  
For example: given an exoplanet in a circumbinary orbit around the ficticious close binary system CT\,Men, one could, in principle, name the exoplanet with any of the following conventions:
CT\,Men\,B, the ``second'' part of the system otherwise consisting of the two stars CT\,Men\,Aa+Ab but potentially containing another stellar system CT\,Men\,C with a totally different dynamical status;
CT\,Men\,C, the third body in the system otherwise consisting of the two stars CT\,Men\,A+B, placing the circumbinary exoplanet on the same hierarchy as the two stars it orbits; or
CT\,Men\,(AB)b, the ``second'' dynamical part of the system otherwise consisting of the two stars CT\,Men\,A+B.
The addition of the form using parentheses to the provisional I.A.U. standard makes it possible to support the last rule.\\
% (e.g. the outer object in a circumbinary system that might otherwise be labeled ``B'' (using the Aa+Ab+B notation) can alternatively be given the name ``(AB)b'' (using the effective convention that A+B=a).
% 4
\item [Rule 4.] {\em When in doubt (i.e. if a different name has not been clearly set in the literature), the hierarchy expressed by the nomenclature should correspond to dynamically distinct (sub-)systems in order of their dynamical relevance.   The choice of hierarchical levels should be made to emphasize dynamical relationships, if known.} \\ [0.1ex]
This rule exploits the implicit freedom within the I.A.U. provisional standard to help decide which hierarchical scheme to adopt.  The examples above clearly show that the new form is the best form for known circumbinary planets and has the nice side-effect of giving these kinds of planets an identical sub-level hierarchical label and stellar component names which conform to the usage within the very close binary community.\\[-2.5ex]
\end{description}
The proposed usage tends to restrict the use of the labels ``C'', ``D'', ... in otherwise undesignated systems, since they are effectively reserved for true trinary \& quartinary systems rather than hierarchical binaries. This is exactly the intent: the names should be as useful as possible.  This restriction could be lifted if the additional capital letters can also be qualified using parentheses: CT\,Men\,(AB)C is a slight variation on the official I.A.U. syntax CT\,Men\,C but which expresses the dynamical information otherwise lost.  The other alternative, CT\,Men\,A(ab)c $\equiv$ CT\,Men\,(ab)c for a system CT\,Men\,A consisting of three components, with component ``c'' orbiting ``a''+``b'', is conceivable but either unwieldy or an implict form compared with the simpler, more explicit, and more familiar-looking alternative CT\,Men\,(AB)b.

Of course, there are situations that can produce unexpected results: imagine a wide binary called WI\,Bin, where PLATO finds single planets around both components as well as a circumbinary planet.   Using the lower-case b notation, these planets would be called WI\,Bin\,b, c, \& d, whereas in the proposed notation they would be WI\,Bin\,Ab, Bb, \& (AB)b since all of them are the first additional objects detected around a higher level object.  
This might seem confusing at first -- all carry the ``b'' suffix -- but this is exactly what is relevant: all of the planets share a common dynamical situation and discovery order.

This nomenclature requires the complete renaming of only two exoplanetary systems, notably HW\,Vir\,3 \& 4 $\rightarrow$ HW\,Vir\,(AB)b \& (AB)c (Lee et al. \cite{lee}; the lower-case b notation is already used in exoplanet.eu) and NN\,Ser\,c \& d  $\rightarrow$ NN\,Ser\,(AB)b \& (AB)c. (Beuermann et al. \cite{beuermann}).
This is no surprise, since these systems are the ones that exposed the naming problems.
The previously known single circumbinary planets PSR\,B1620-26\,b (Thorsett, Arzoumanian \& Taylor \cite{PSR}) and DP\,Leo\,b (Qian et al. \cite{qian}) can almost retain their names as unofficial informal forms of the ``(AB)b'' designation where the ``(AB)'' is left out.

Note that this slightly revised naming convention follows the present provisional I.A.U. convention in that it does not distinguish between stars, brown dwarfs, or planets.
Since many quoted exoplanet masses are actually lower limits, this is probably just as well.

%______________________________________________________________

\section{Conclusions}

We have proposed a slight revision of the provisional I.A.U. nomenclature standard for multiple systems with the intent of reaching an effective and simple naming convention for extrasolar planets.
Our proposal is nearly 100\% compatible with both the standard and common exoplanet usage and yet permits one to distinguish clearly between the dynamical status of planets around single stars, stars in multiple systems, and circumbinary (or higher order) planets.
While primarily designed for the exoplanet community, the parenthesis syntax could naturally be used to good effect for stellar multiple systems as well.
Thus, we encourage a broadly based discussion on the feasibility of endorsing this (or a similar) convention in the hopes of giving our objects useful and uniform names.

\begin{figure}
\begin{center}
\includegraphics[width=72.0mm]{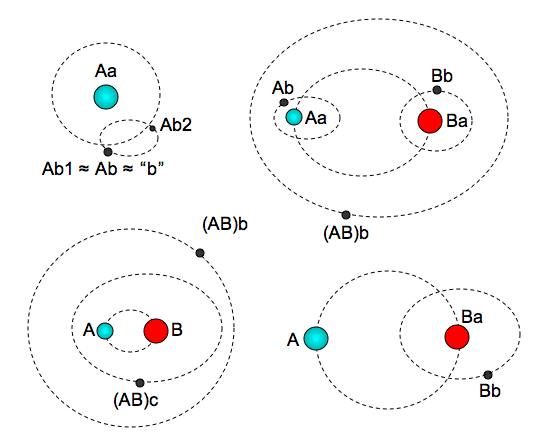}
\end{center}
\caption{Examples of different exoplanet name suffixes in single and binary systems using the proposed system. Upper left: exoplanet around a single star (e.g. 51\,Peg) plus a moon.  Upper right: double star, each with a planet (e.g. HD\,41004), plus a circumbinary planet.  Lower left: two circumbinary planets (e.g. NN\,Ser).  Lower right: planet around the secondary star in a binary (e.g. HD\,178911).}
\end{figure}

%______________________________________________________________

\begin{acknowledgements}
We would like to acknowledge the use of the exoplanet.eu online database and would like to thank the referees of the Beuermann et al. paper for bringing up this issue for the case of the planets tentatively named NN\,Ser\,c and NN\,Ser\,d (i.e. NN\,Ser\,(AB)b and (AB)c in our proposed notation) and for informal discussions with other colleagues and members of the I.A.U. Commission 53.
This letter was originally submitted to A\&A with the intent of reaching as wide an audience as possible.
We would like to thank the editors and referees for their very positive comments, even though they ultimately decided to reject this letter on the grounds that publication might make this a de-facto standard independent of any action by the I.A.U.
\end{acknowledgements}

\end{document}